\documentstyle[twocolumn,aps,epsf]{revtex}
\begin{document}
\draft
\title{Excitonic Dynamical Franz-Keldysh Effect}
\author
{K. B. Nordstrom$^1$, K. Johnsen$^{2,3}$, S. J. Allen$^1$, A.-P.
Jauho$^2$,
B. Birnir$^3$, J. Kono$^4$,T. Noda$^5$,\\
H. Akiyama$^5$, and H. Sakaki$^{4,5}$}
\address{
$^1$Center for Terahertz Science and Technology, Quantum Institute,
University of California,\\
Santa Barbara, California, 93106, USA\\
$^2$Mikroelektronik Centret, Technical University of Denmark, Bldg
345east\\
DK-2800 Lyngby, Denmark\\
$^3$Mathematics Department, University of California, Santa Barbara,
California, 93106, USA\\
$^4$Quantum Transition Project, Japan Sci. \& Tech. Corp., Tokyo 153,
Japan\\
$^5$Institute of Industrial Science, University of Tokyo, Tokyo 106,
Japan\\
\medskip
\date{Draft: \today}
\parbox{14cm}{\rm
The Dynamical Franz-Keldysh Effect is exposed by exploring
near-bandgap absorption in the presence of intense THz
electric fields.  It bridges the gap between the DC Franz-
Keldysh effect and multi-photon absorption and competes with
the THz AC Stark Effect in shifting the energy of the
excitonic resonance.  A theoretical model which includes the
strong THz field non-perturbatively via a non-equilibrium
Green Functions technique is able to describe the Dynamical
Franz-Keldysh Effect in the presence of excitonic absorption.
\smallskip\\}
}
\maketitle
\narrowtext

We report observation of the Dynamical Franz Keldysh Effect (DFKE)
by examination of the near-bandgap optical properties of a semiconductor
multiple quantum well (MQW) in a new experimental regime. By applying a
strong electric field at frequencies near 1 THz, we perform
electro-absorption spectroscopy which is described by neither the DC
Franz-Keldysh Effect (FKE) \cite{franz,keldysh}, nor by
optical-frequency effects like multi-photon absorption (MPA) \cite{MPA}.
The DFKE region, bridging these two extremes, is a topic of
longstanding theoretical interest \cite{yacoby,b3,REB85}, but has not
previously been studied experimentally.

The crossover between ``high''
and ``low'' frequency may be quantified in terms of two parameters,
$E_{\rm KE}$ and $\gamma$, defined as:
\begin{equation}
E_{\rm KE} = {e^2 E_{\rm{THz}}^2 \over 4 m^* \omega_{\rm{THz}}^2}.
\end{equation}
\begin{equation}
\gamma = {E_{\rm KE} \over \hbar \omega_{\rm{THz}}} =
{e^2 E_{\rm{THz}}^2 \over 4 m^* \hbar \omega_{\rm{THz}}^3}.
\end{equation}
$E_{\rm KE}$ is the mean kinetic
energy of a particle of mass $m^*$ and charge $e$ in an electric field
$\vec E_{\rm{THz}}\cos (\omega_{\rm{THz}} t)$,
{\it i.e.} the ponderomotive energy.
The DC FKE corresponds to $\gamma \gg 1$;
multi-photon effects occur for $\gamma\ll 1$;
and the case of $\gamma \sim 1$ is a new, distinct DFKE
regime \cite{yacoby} (Fig. 1).  This regime is difficult to access
experimentally, as $E_{\rm KE}$ must be comparable to or greater than
any broadening energies in the system.  However, it is possible for
applied fields with $\omega_{\rm{THz}} \sim 1$THz and
$E_{\rm{THz}} \sim\rm{1-10kV/cm}$ \cite{b3}.

In this paper, we explore the optical absorption of a semiconductor
multiple quantum well (MQW) driven by an intense THz electric field
polarized in the plane of the MQW layers.  For the case of
non-interacting carriers, the DFKE predicts a blue-shift of the main
absorption edge in the spectrum by $E_{\rm KE}$, as well as increased
sub-gap absorption\cite{b3} (Fig. 2a).  This DFKE blue-shift has
not previously been observed.

In order to properly address the DFKE, we must also
consider the effect of excitons, which dominate the observed
near-bandgap absorption in undoped semiconductors.
Excitons are bound states of an electron and hole with hydrogen-like
quantized energy levels.  The energy level spacings can be in the THz
regime.
The interaction of excitonic states with a THz
field leads to shifts of the levels via the AC Stark
Effect\cite{stark} (Fig. 2b). The THz field interacts most strongly with
the exciton $1s \rightarrow 2p$ transition\cite{junprl}.
We present below a novel theoretical model which shows that the
experimentally
observed changes in the absorption spectrum, as a function of the THz
field intensity and frequency, display the DFKE and its
interplay with the AC Stark Effect.

Our MBE-grown sample consists of a 20-period MQW with 80\AA\ In$_{\rm
0.2}$Ga$_{\rm 0.8}$As wells separated by 150\AA\ GaAs barriers on a GaAs
substrate. The energy of the 2D MQW states is below the substrate
band-gap, allowing us to perform direct transmission spectroscopy
without substrate removal.  Strain lifts the valence-band degeneracy at
$k=0$, splitting the light-hole and heavy-hole bands by 50meV.  The
electron-heavy-hole exciton is the lowest-energy feature.  It is
inhomogeneously broadened by interface roughness and alloy
disorder\cite{ingaas} with a FWHM of 6meV. The exciton $1s \rightarrow
2p$ transition energy ($\hbar\omega_{12}$) is $\sim$8meV, estimated from
magneto-transmission experiments\cite{junprl}.

A CW Ti-Sapphire laser provides tunable near-bandgap radiation,
in the near infrared (NIR),
focussed to a $\sim 200\mu$m spot.  Its transmission is detected by a
conventional Si PIN photo-diode, and peak NIR intensity is $\sim100{\rm
mW}/{\rm cm}^2$. THz radiation is provided by the UCSB Free-Electron
Laser, a source of intense, coherent radiation tunable from
$\hbar\omega_{\rm THz}\sim$0.5-20meV\cite{jerry}.
It is roughly collinear with the NIR beam
and focussed to a 0.5-2.5mm spot coincident with the NIR focus (Fig.
2c).  We measure transmission spectra as a function of THz intensity and
frequency.  All measurements are performed in He vapor at $\sim$7K.

We observe two distinct regimes of experimental behavior for different
ranges of THz frequency, $\omega_{\rm{THz}}$.  When $\omega_{\rm{THz}}$
is less than $\omega_{12}$, we observe a red-shift of the exciton for
low THz intensities, \textit{i.e.}, an AC Stark shift. As THz intensity
increases, this shift reaches a maximum and reverses, eventually
becoming a net blue-shift at the highest THz intensity; the DFKE
blue-shift dominates at high intensities as the AC Stark Effect
saturates (Fig. 3a). Conversely, if $\omega_{\rm{THz}}$ is greater than
$\omega_{12}$, we observe a blue-shift of the exciton which increases
monotonically with increasing THz intensity; the AC Stark Effect and
DFKE act in concert, each contributing to the blue-shift (Fig. 3b). In
both cases, the exciton peak is broadened and suppressed, and the
broadening increases with increased THz intensity (Fig. 3a, b).

Since the apparent peak shifts are small, we quantify them by
taking a weighted average of the data near the exciton line.  We find
the geometric mean of the upper half of the excitonic peak, and plot
this ``center of mass'' against $\gamma$, which is directly proportional
to THz power (Fig. 4).  Note that our estimates of the maximum
$\gamma$ at each frequency are only accurate to within a factor of 2 or
3.
We also show results of our theoretical calculations (to be discussed
below).
Again, we see an initial red-shift, followed by a blue-shift at low
$\omega_{\rm{THz}}$ (Fig. 4a) and a monotonic blue-shift at high
$\omega_{\rm{THz}}$ (Fig. 4b).

Theoretically, we describe the optical properties of the system by
calculating the time dependent induced polarization. The
system is probed using a weak NIR probe field
${\vec E}_{\rm NIR}(t)$ with frequency $\omega \sim E_{\rm gap}/\hbar$.
To linear order in the NIR field, the macroscopic polarization is the
$\vec k$-trace of
\begin{equation}
{\vec P}(\vec k,t) = \int_{-\infty}^\infty{\rm d}t'
\,\chi^r (\vec k;t,t'){\vec E}_{\rm NIR}(t').
\label{ad1}
\end{equation}
All the information about the THz field is contained in the
non-equilibrium generalization of the retarded two-time interband
susceptibility $\chi^r (\vec k;t,t')$.
In our previous work \cite{b3},
based on non-equilibrium Green functions, we studied the
susceptibility starting from a two-band Hamiltonian,
including the THz field non-perturbatively
in the Coulomb-gauge via the vector potential
$\vec A(t) = -\vec E_{\rm THz}\sin (\omega_{\rm THz} t)/\omega_{\rm
THz}$.
Here, $\vec E_{\rm THz}$
is oriented in the plane of the MQW layers.
Non-equilibrium Green function
techniques \cite{b4} yield an analytic expression for
the single particle susceptibility:
\begin{equation}
\bar\chi^r(\vec k;t,t') =  -{2d^2\over\hbar}\theta (t-t')
\sin \Big\{
\int_{t'}^t{{\rm d}s\over\hbar} \epsilon[\hbar\vec k-e\vec A(s)]
\Big\},
\label{chigen}
\end{equation}
where $d$ is a dipole matrix-element and $\epsilon [\vec k] = { k^2
\over 2
\mu} + E_{\rm gap}$, $\mu$ being the electron-hole reduced mass and
$\vec k$
the in-plane momentum.

We include excitonic effects by using
the Bethe-Salpeter equation \cite{b2} to determine a
susceptibility which includes the electron-hole interaction in the
ladder approximation:
\begin{eqnarray}
&&\chi^r (\vec k;t ,t' ) = \bar\chi^r (\vec k;t,t')
\nonumber \\&&+\int {{\rm d}^2\vec k'\over (2\pi )^2}
\int_{-\infty}^\infty{\rm d}t''\; \bar\chi^r (\vec k;t,t'' )
V(|\vec k - \vec k'|) \chi^r (\vec k';t'',t' ),
\label{inteq}
\end{eqnarray}
where $\bar\chi^r (\vec k;t,t')$ is the single-particle, non-interacting
susceptibility of Eq. (\ref{chigen}).
Expressing the susceptibility as
$\chi^r (\vec k;t ,t' ) =
\sum_n\int{{\rm d}\omega\over 2\pi}\;\chi_n^r(\vec k,\omega)
e^{i\omega (t-t') + in2\omega_{\rm THz} (t+t')}$,
the integral equation (\ref{inteq}) becomes a
matrix equation \cite{b5},
\begin{eqnarray}
&&\chi_n^r(\vec k,\omega) = \bar\chi_n^r(\vec
k,\omega)
\\\nonumber&&+\sum_{n'}\bar\chi_{n-n'}^r(\vec
k,\omega+2n'\omega_{\rm THz})
\\\nonumber &&\times\int {{\rm d}^2\vec k'\over (2\pi )^2}
V(|\vec k - \vec k'|)\chi_{n'}^r(\vec k',\omega+2(n'-n)\omega_{\rm
THz}).
\end{eqnarray}
For physically achievable THz intensities, only small values of $n$ are
needed.

Numerically, we solve the matrix equation including both
$s$-wave and $p$-wave scattering; both are important, since the THz
field
will couple the $s$ and $p$ states of the exciton and influence the
observed resonance via the AC Stark Effect. The solutions to the
resulting equation are found by discretizing the integrals to yield a
set of linear equations which we solve by standard methods \cite{b6}.

Finally, we relate the results to physically measurable quantities by
expressing the macroscopic polarization as
\begin{equation}
\vec P(t) =
\vec E_{\rm NIR}
\sum_n \tilde\chi_n(\omega ) e^{i(2n\omega_{\rm THz}-\omega )t},
\end{equation}
where $\tilde\chi_n(\omega )=\sum_{\vec k} \chi_n^r(\vec k,\omega )$.
The linear absorption is thus proportional to Im$\tilde\chi_0(\omega )$.
The terms with $n\neq 0$ describe the non-linear mixing of the NIR and
the
THz field, resulting in optical sideband generation.
These sidebands show a rich structure, but an analysis is beyond the
scope of this work, and will be reported separately.

Using the above method, we have calculated absorption spectra using the
experimental parameters of our system.  We find good agreement
with experiment, using no fitted parameters.
For $\omega_{\rm{THz}} < \omega_{12}$, theory predicts a red-shift of
the exciton absorption peak at low THz intensities.  With increasing THz
intensity, the shift saturates and then reverses, eventually becoming a
net blue-shift (Fig. 5a).  Theory also predicts a blue-shift for
$\omega_{\rm{THz}} > \omega_{12}$, which monotonically increases with
THz intensity for experimentally-accessible intensities (Fig. 5b).  In
both cases, the exciton peak is suppressed and inhomogeneously broadened
with increasing intensity, \textit{i.e.}, with increasing $\gamma$ (Fig.
5ab).  Using the same method as for
the measured spectra, we compute the ``center of mass'' of the
calculated
exciton lines.  We plot the shift of the theoretical ``center of mass''
as
a function of $\gamma$ in Fig. 4, along with the experimental results.
The theoretical and experimental
peak shifts show identical qualitative behavior.

As described above, this behavior can be understood as the DFKE acting
in competition with the AC Stark Effect.  The AC Stark Effect, at
low $E_{\rm{THz}}$, results in a
shift of the exciton level,
\begin{equation}
\Delta \propto  {(\omega_{\rm{THz}} - \omega_{12})
E_{\rm{THz}}^2
\over
(\omega_{\rm{THz}} - \omega_{12})^2 + \Gamma^2},
\end{equation}
where $\Gamma$ is the width of the $1s \rightarrow 2p$ transition
line \cite{stark}.
We estimate $\hbar\Gamma \sim 4$meV in our sample.  The magnitude of
$\Delta$
is dominated by $\Gamma$, and we expect little resonant enhancement of
$|\Delta|$ for $\omega_{\rm{THz}} \sim \omega_{12}$.
$\Delta$ does not increase indefinitely, and will
saturate as $E_{\rm{THz}}$ increases,
going from quadratic to linear
dependence on $E_{\rm THz}$ \cite{stark}.
The DFKE, however, always provides a
blue-shift proportional to $E_{\rm{THz}}^2$, given by $E_{\rm KE}$
\cite{b3}.

The net effect is illustrated in figure 4, which shows the shift of the
center of mass of the exciton transmission peak as a function of
$\gamma$ for both theory and experiment.
For $\omega_{\rm{THz}} < \omega_{12}$,
we see a red shift which saturates and reverses with increasing
intensity, showing a roughly linear dependence on THz intensity at high
fields (Fig. 4a).  At low fields, $\Delta$ dominates, resulting in a net
red shift.  As $\Delta$ saturates with increasing field, $E_{\rm KE}$
begins to dominate, eventually overwhelming the red shift entirely and
resulting in a net blue shift. For $\omega_{\rm{THz}} > \omega_{12}$, we
observe only an increasing blue-shift with increasing $E_{\rm{THz}}$
(Fig. 4b). Here, $\Delta$ and $E_{\rm KE}$ cooperate to create a
blue-shift.

In conclusion, we have observed the experimental signature of DFKE
through its interplay with the AC Stark Effect.  Our theory and
observations are in remarkable agreement, and show the resulting
frequency- and intensity-dependent shift of the excitonic resonance in a
MQW
under intense THz irradiation.

This work is supported by ONR grant N00014-92-J-1452, the Quantum
Transition Project of JSTC, QUEST (an NSF Science \& Technology Center),
LACOR grant no. 4157U0015-3A from Los Alamos National Laboratory,
National Science Foundation grant CDA96-01954, and by Silicon Graphics,
Inc.  The authors also wish to thank D.P. Enyeart, D.T. White, J.R.
Allen and G. Ramian at the Center for Terahertz Science \& Technology
for their invaluable technical support.

\begin{figure}
\epsfxsize=8.5cm\epsfbox{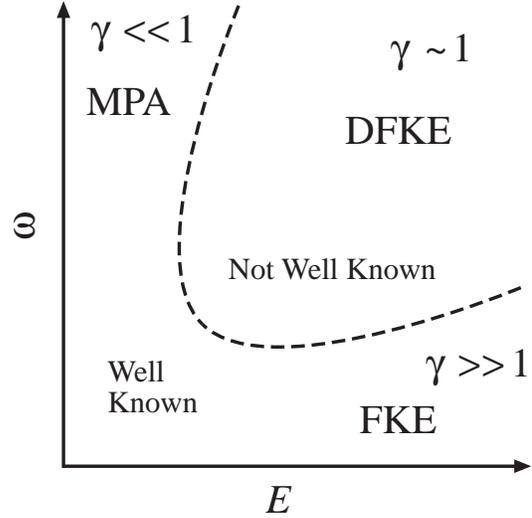}
\caption{
A schematic map of electro-optic phase space as a function of amplitude
and
frequency of the perturbing field, showing DFKE regime ``between'' MPA
and FKE.
}
\label{fig1}
\end{figure}
\begin{figure}
\epsfxsize=8.5cm\mbox{\hspace{0.1cm}\epsfbox{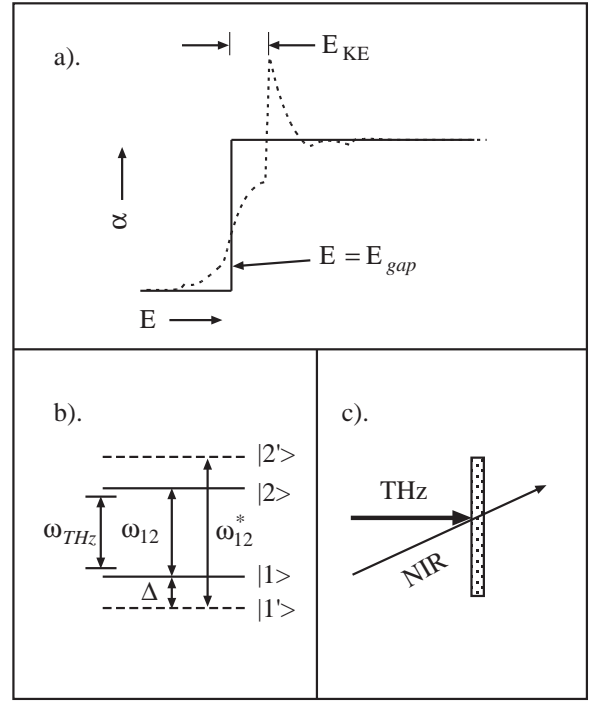}}
\caption{
(a) DFKE in an ideal 2D system.  In a strong THz field (dashed curve,
solid
curve is zero-field), the ``edge'' shifts to higher frequency and
sub-gap
absorption increases.  (b)  The AC Stark Effect: a strong field applied
at
frequency $\omega \sim\omega_{12}$ causes $\omega_{12}$ to shift.  For
$\omega <(>) \omega_{12}$, the transition shifts to
$\omega_{12}^* >(<) \omega_{12}$.  $\Delta = (\omega_{12} -
\omega_{12}^*)/2$ is the shift of $|1\rangle$ with respect to distant
energy levels.  (c) Schematic of experiment.
}
\label{fig2}
\end{figure}
\begin{figure}
\epsfxsize=8.5cm\mbox{\hspace{0.1cm}\epsfbox{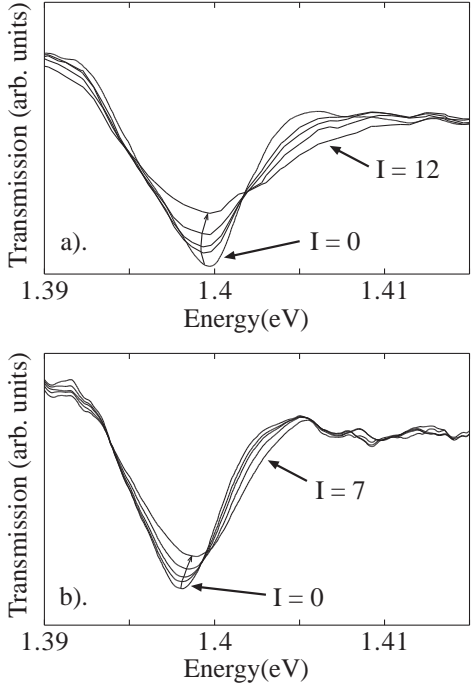}}
\caption{
Experimental transmission of MQW near e1h1 exciton with
(a) $\hbar \omega_{\rm THz} = 2.5$meV at $I_{\rm THz} = 0, 1, 2,
4, 12$ (arb. units).
(b) $\hbar \omega_{\rm THz} = 14$meV at $I_{\rm THz} = 0, 1, 2,
4, 7$ (arb. units).
Arrows connect calculated centers of experimental peaks
and point in direction of increasing $I_{\rm THz}$.
}
\label{fig3}
\end{figure}
\begin{figure}
\epsfxsize=8.5cm\mbox{\hspace{0.1cm}\epsfbox{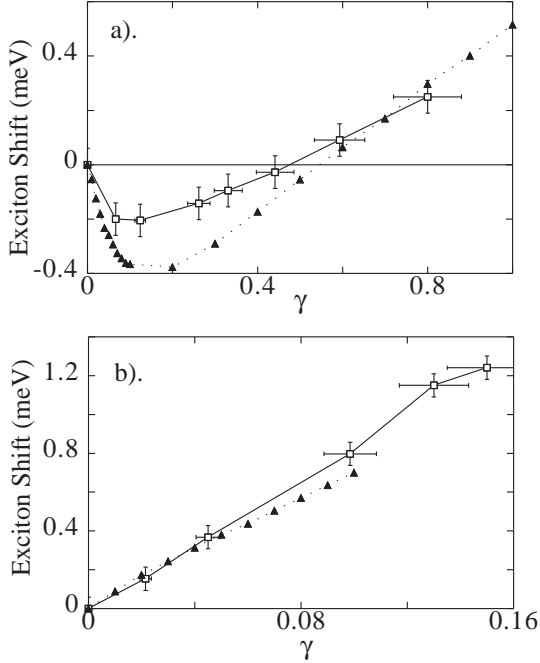}}
\caption{
``Center of mass'' of measured (squares) and calculated (triangles)
exciton transmission peak vs. $\gamma$ for
(a) $\hbar\omega_{\rm THz} = 2.5{\rm meV}
< \hbar\omega_{12}$ (b) and
$\hbar\omega_{\rm THz} =
10.5{\rm meV} > \hbar\omega_{12}$.
}
\label{fig4}
\end{figure}
\begin{figure}
\epsfxsize=8.5cm\begin{center}\mbox{\epsfbox{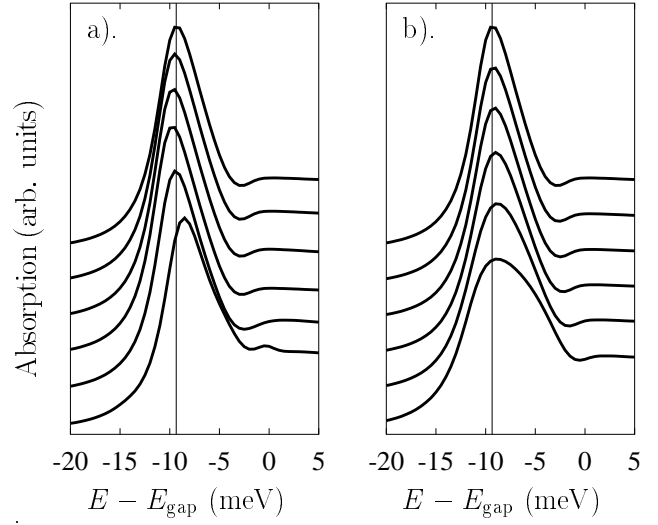}}\end{center}
\caption{
Calculated absorption of MQW with; (a)
$\hbar \omega_{\rm THz} = 2.5$meV $< \hbar\omega_{12}$ from top down at
$\gamma =$ 0, 0.03, 0.06, 0.1, 0.5, 1.5.
(b) $\hbar \omega_{\rm THz} = 14$meV $> \hbar\omega_{12}$
from top down at $\gamma =$ 0, 0.01, 0.02, 0.04, 0.08, 0.15.
}
\label{fig5}
\end{figure}

\end{document}